\documentclass[preprint,aps,prd,superscriptaddress,longbibliography]{revtex4-1}

\usepackage{bm}
\usepackage{graphicx}
\usepackage{booktabs}
\usepackage{slashed}
\usepackage{hyperref}
\usepackage{amssymb,amsmath,amsfonts}
\usepackage{color}
\usepackage[utf8]{inputenc}
\usepackage[caption=false]{subfig}
\newcommand{\sqz}{{\cal Z}}
\newcommand{\meff}{m_{{\rm eff}}}

\newcommand{\be}{\begin{equation}}
\newcommand{\ee}{\end{equation}}
\newcommand{\bea}{\begin{eqnarray}}
\newcommand{\eea}{\end{eqnarray}}

\newcommand{\p}{\partial}
\newcommand{\s}{\sigma}

\newcommand{\la}{\langle}
\newcommand{\ra}{\rangle}
\newcommand{\rd}{\mbox{d}}
\newcommand{\ri}{\mbox{i}}
\newcommand{\re}{\mbox{e}}

\renewcommand{\vec}[1]{{\bm #1}}

\begin{document}

\title{Fluctuations in cool quark matter and the phase diagram of Quantum Chromodynamics}

\author{Robert D. Pisarski}
\affiliation{Department of Physics, Brookhaven National Laboratory, Upton, NY 11973}

\author{Vladimir V. Skokov}
\affiliation{RIKEN/BNL Research Center, Brookhaven National Laboratory, Upton, NY 11973}
  
\author{Alexei M. Tsvelik}
\affiliation{Condensed Matter Physics and Materials Science Division, Brookhaven National Laboratory, Upton, NY 11973-5000, USA}

\begin{abstract}

We consider the phase diagram of hadronic matter as a function of
temperature, $T$, and baryon chemical potential, $\mu$.
Currently the dominant paradigm 
is a line of first order transitions which ends at a critical endpoint.

In this work we suggest that spatially inhomogenous phases are a generic
feature of the hadronic phase diagram at nonzero $\mu$ and low $T$.
Familiar examples are pion and kaon condensates.
At higher densities, we argue that these condensates connect onto
chiral spirals in a quarkyonic regime.
Both of these phases exhibit the spontaneous breaking of a global $U(1)$ symmetry
and quasi-long range order, analogous to smectic liquid crystals.
We argue that there is a continuous line of first order transitions which separate
spatially inhomogenous from homogenous phases, where the latter can be either a hadronic phase or a quark-gluon plasma.

While mean field theory predicts that there is a Lifshitz point along this line of first order transitions, 
in three spatial dimensions strong infrared fluctuations wash out any Lifshitz point.
Using known results from inhomogenous polymers, we suggest that instead there is 
a Lifshitz regime.  Non-perturbative effects are large in this regime, where
the momentum dependent terms
for the propagators of pions and associated modes
are dominated not by terms quadratic in momenta, but quartic.
Fluctuations in a Lifshitz regime may be directly relevant to the collisions
of heavy ions at (relatively) low energies, $\sqrt{s}/A: 1 \rightarrow 20$~GeV.

\end{abstract}

\maketitle

\section{Introduction}

The phases of Quantum Chromodynamics (QCD),
as a function of temperature, $T$, and the baryon (or equivalently, quark)
chemical potential, $\mu$, are of fundamental interest \cite{Kogut:2004su, *Yagi:2005yb, *Fukushima:2010bq}.
At zero chemical potential, numerical simulations on the lattice indicate that there is
no true phase transition, just a crossover, albeit one where the degrees of freedom increase
dramatically
\cite{borsanyi_full_2014,bazavov_equation_2014,ratti_lattice_2016}.
At a nonzero chemical potential, however, 
a crossover line may meet a line of first order transitions at a critical endpoint
\cite{Asakawa:1989bq,Stephanov:1998dy,Stephanov:1999zu,Son:2004iv,Hatta:2002sj,Stephanov:2008qz}.
These lines separate two phases: a hadronic phase in which chiral symmetry is spontaneously
broken, and a (nearly) chirally symmetric phase of quarks and gluons.

In condensed matter it is well known that a third phase can arise,
in which spatially inhomogeneous structures form.
If so, the three phases meet at a Lifshitz point
\cite{Hornreich:1975zz,Hornreich:1980,chaikin:2010,Diehl:2002a,Erzan:1977,Sak:1978,Grest:1978,Fredrickson:1997,Duchs:2003,Fredrickson:2006,JonesLodge,Cates12,Bonanno:2014yia,Zappala:2017vjf,Zappala:2018khg}.

In hadronic nuclear matter the existence of spatially inhomogenous phases is familiar, as pionic
\cite{Overhauser:1960,Migdal:1971,Sawyer:1972cq,Scalapino:1972fu,Sawyer:1973fv,Migdal:1973zm,Migdal:1978az,Migdal:1990vm,Kleinert:1981zv,Baym:1982ca,Kolehmainen:1982jn,Bunatian:1983dq,Takatsuka:1987tn,Kleinert:1999gc}
and kaonic \cite{Kaplan:1986yq,Brown:1993yv,Brown:1995ta,Brown:2007ara} condensates.
They are ubiquitous in Gross-Neveu models in $1+1$ dimensions, which are soluble either for a large number
of flavors \cite{Schon:2000he,Schnetz:2004vr,Thies:2006ti,Basar:2008im,Basar:2008ki,Basar:2009fg} or by using advanced nonperturbative techniques \cite{Azaria:2016}.
These phases also arise from analyses of effective models of QCD, where they have been
termed chiral spirals
\cite{Deryagin:1992rw,Shuster:1999tn,Park:1999bz,Rapp:2000zd,Nakano:2004cd,Bringoltz:2006pz,Sadzikowski:2006jq,Bringoltz:2008iu,Miura:2008gd,Bringoltz:2008iu,Bringoltz:2009ym,Maedan:2009yi,Nickel:2009ke,Nickel:2009wj,Abuki:2010jq,Carignano:2010ac,Partyka:2010em,Carignano:2012sx,Buballa:2012vm,Kamikado:2012cp,Abuki:2013pla,Feng:2013tqa,Karasawa:2013zsa,Moreira:2013ura,Muller:2013tya,Carignano:2014jla,Hayata:2014eha,Kitazawa:2014sga,Kojo:2014vja,Kojo:2014fxa,Braun:2015fva,Buballa:2015awa,Carignano:2015kda,Carlomagno:2015nsa,Harada:2015lma,Adhikari:2016vuu,Adhikari:2016jzc,Carignano:2016jnw,Ferrer:2016osp,Karasawa:2016bok,Yokota:2016tip,Khunjua:2017khh,Carignano:2017meb,Heinz:2013hza,Heinz:2015lua,Buballa:2014tba,Tatsumi:2014cea,Lee:2015bva,Hidaka:2015xza,Nitta:2017mgk,Nitta:2017yuf,Yoshiike:2017kbx,McLerran:2007qj,Kojo:2009ha,Kojo:2010fe,Kojo:2011cn,Andronic:2009gj}.
In this paper we consider especially the role played by fluctuations in phases with spatially
inhomogeneous phases, and show that they can dramatically affect the phase diagram of QCD.

At densities between a dense hadronic phase and deconfined quarks,
cool quark matter is quarkyonic: while the pressure is (approximately) perturbative,
the excitations near the Fermi surface are confined \cite{McLerran:2007qj}.
The Fermi surface of quarks, which starts out as isotropic, breaks up into a set of patches,
with the longitudinal fluctuations governed by a Wess-Zumino-Novikov-Witten
Lagrangian \cite{Kojo:2009ha,Kojo:2010fe,Kojo:2011cn}.
The transverse fluctuations, however, are of higher order in momenta: they are not quadratic,
but quartic.   Among condensed matter systems the closest analogy are  smectic 
liquid crystals, which consist of elongated molecules periodically ordered in just one direction \cite{chaikin:2010}.
The color and flavor quantum numbers which quarks carry makes the analogous state more involved.
As for pion/kaon condensates 
\cite{Kleinert:1981zv,Baym:1982ca,Kolehmainen:1982jn,Bunatian:1983dq,Takatsuka:1987tn,Kleinert:1999gc,Lee:2015bva,Hidaka:2015xza,Nitta:2017mgk,Nitta:2017yuf}, 
at nonzero temperature there are long range correlations in the inhomogeneous phase,
and there is no true order parameter.  At nonzero temperature in $3+1$ dimensions,
the fluctuations exhibit complicated patterns.
The propagators of two-quark operators decay exponentially with a temperature dependent correlation length,
and propagators of spin and flavor singlet operators, composed of $2N_f$ quarks, fall off as a power law.

We then consider how a quarkyonic phase matches onto the usual hadronic phase at low temperature
and density.  We first discuss how as the density decreases, the chiral spirals in a quarkyonic phase transform naturally
into the pion/kaon condensates of hadronic nuclear matter.
We show that the phase diagram valid in
mean field theory --- two lines of second order phase transitions, meeting a line of first order
transitions at a Lifshitz point --- is dramatically altered by fluctuations.  As demonstrated first
by Brazovskii \cite{Brazovski:1975,Dyugaev:1982gf,Hohenberg:1995,Karasawa:2016bok}, the line
of transitions between the symmetric phase and that with chiral spirals becomes a line
of first order transitions.  Most importantly,
the infrared fluctuations about a Lifshitz point are so strong that there is, in fact,
no true Lifshitz point
\cite{Erzan:1977,Sak:1978,Grest:1978,Bonanno:2014yia,Zappala:2017vjf,Zappala:2018khg}. This is known
to occur in inhomogenous polymers, both from experiment and numerical simulations
\cite{Fredrickson:1997,Duchs:2003,Fredrickson:2006,JonesLodge,Cates12}.
We suggest that in QCD, infrared fluctuations also wipe out the Lifshitz point, 
leaving just a line of first order transitions separating the region with
inhomogeneous phases from those without.  What remains is a Lifshitz regime,
which is illustrated in Figs. \ref{fig:fluc_phase} and \ref{fig:qcd} below.
The Lifshitz regime is manifestly non-perturbative, as the momentum dependence
for the propagators for pions (or the associated modes of a chiral spiral) are dominated
by terms quartic, instead of quadratic, in the momenta.
This change in the momentum dependence generates large fluctuations,
which may be related to
known
\cite{Andronic:2009gj,Andronic:2014zha,Andronic:2017pug,Adamczyk:2017iwn,Bugaev:2017nir,Bugaev:2017bdc,Bugaev:2018tyt}
and possible \cite{Jowzaee:2017ytp,Luo:2017faz}
anomalies in the collisions of heavy ions at center of mass energies per nucleon $\sqrt{s}/A: 1 \rightarrow 20$~GeV.

\section{ The model of quarkyonic phase}
\label{model_quarkyonic}

A quarkyonic phase only exists when quark excitations near the Fermi surface are (effectively) confined.
  While for three colors numerical simulations of lattice gauge theories at nonzero quark density
  are afflicted by the sign problem for
  three colors, they are possible for two colors
  \cite{Hands:2010gd,Hands:2011ye,Cotter:2012mb,Braguta:2015owi,Braguta:2016cpw,Bornyakov:2017txe}.
  Although the original argument for a quarkyonic phase was based upon
  the limit of a large number of colors \cite{McLerran:2007qj}, these simulations show that
  even for two colors, the expectation value of the Polyakov
  loop is small, indicating confinment, up to large values of the quark
  chemical potential   \cite{Hands:2010gd,Hands:2011ye,Cotter:2012mb,Braguta:2015owi,Braguta:2016cpw,Bornyakov:2017txe}.
  Directly relevant to our analysis are the results of Bornyakov {\it et al.}, who find that the string tension
  decreases gradually from its value in vacuum, to essentially zero at
  $\mu_q \sim 750$~MeV: see Fig. 3 of Ref.~\cite{Bornyakov:2017txe}.
  This suggests that a quarkyonic phase may dominate for a wide range of chemical potential, and indeed,
  for all values relevant to hadronic stars \cite{Kurkela:2014vha,Fraga:2015xha}.

  There is an elementary argument for why there could be such a large region in which cool quark matter
  is quarkyonic.  Consider computing a scattering process in vacuum.   By asymptotic freedom, this is certainly valid
  at large momentum.  As the momentum decreases, non-perturbative effects enter, and a perturbative computation
  is invalid.  This certainly occurs by momenta $\Lambda_{{\rm scat}} \sim 1$~GeV, if not before.

  Now consider computing the pressure perturbatively.  The scattering processes which contribute involve
  the scattering of quarks and holes with momenta whose magnitude is on the order of the Fermi momentum.
  This suggests that the pressure can be computed perturbatively down to a scale which is typical of
  that where perturbative computations are valid: that is, on the order of $\Lambda_{{\rm scat}} \sim 1$~GeV.
  This argument is clearly qualitative: the estimate of $\Lambda_{{\rm scat}}$ as applying to the perturbative
  computation of the pressure could well vary by a factor of two.
  Furthermore, our argument applies only to the pressure: excitations about the Fermi surface involve
  much smaller energies than the chemical potential, so that one expects a transition from the
  perturbative, to the quarkyonic, regime \cite{McLerran:2007qj}.
  What is important is that the momentum scale $\Lambda_{{\rm scat}}$ does not depend
  strongly upon the number of colors.  
  This elementary argument may explain why there is a large quarkyonic regime even for two colors
  \cite{Hands:2010gd,Hands:2011ye,Cotter:2012mb,Braguta:2015owi,Braguta:2016cpw,Bornyakov:2017txe}.

  This estimate differs from that at zero quark density and a nonzero temperature, $T$.
  In computing the pressure at $T \neq 0$ and $\mu_q = 0$, the dominant
  momentum scale is naively the first Matsubara frequency,
  $= 2 \pi T$ \cite{Braaten:1995ju,Braaten:1995jr}.  Detailed computations
  to two loop order \cite{Laine:2005ai,Schroder:2005zd} show that the precise value is a bit larger,
  but this estimate is qualitatively correct.  At temperatures $\sim 150$~MeV, this is about $\sim 1$~GeV,
  which is the same as we estimate for $T = 0$ and $\mu_q \neq 0$.

  We note that a quantitative measure of what momentum scale perturbative computations of the pressure
  are valid will be provided by computations to high order, such as $\sim g^6$, as are currently
  underway \cite{Kurkela:2016was,Ghisoiu:2016swa}.

  Returning to the quarkyonic phase, previously we assumed that the chiral symmetry was restored
  \cite{Kojo:2009ha,Kojo:2010fe,Kojo:2011cn}.
  However, for the reasons of convenience we will work with the
  nonrelativistic limit of the quark Hamiltonian which is formally justified if the 
the constituent quark mass is nonzero. 
At nonzero quark mass, we work with the nonrelativistic limit of the quark Hamiltonian:

\bea
H &=& \int \rd^3x \Big\{\psi^+_{\alpha,f,a}\Big[ \frac{1}{2m_f}\left(\ri{\bf\nabla}
  - \frac{q_f}{c}{\bf A}\right)^2 + (q_f/m_fc^2)(\vec\s[{\bf \nabla}\times{\bf A}])
- (\mu - m_fc^2)\Big]\psi_{\alpha,f,a} \nonumber\\
&& + V_1 + V_2\Big\} \; ,
\label{H}\\
V_1 &=& \frac{1}{2} \; \left( \psi^+(r) T^A\psi(r)\right)
D_{00}(|{\bf r}-{\bf r}'|)
\left( \psi^+(r')T^A\psi(r') \right) \; ,
\label{V1}\\
V_2 &=& \frac{1}{2} \; \left( \psi^+(r)\vec\s\otimes T^A\psi(r) \right)
D_{\perp}(r-r')
\left(\psi^+(r')\vec\s \otimes T^A\psi(r') \right) \; ,
\label{V2}
\eea
%
where $\psi^+T^A\psi = \left( \psi^+_{\alpha,f;a}(r) T^A_{a b} \psi_{\alpha,f;b}(r) \right)$ and $\psi^+\otimes \s^aT^A\psi = \left( \psi^+_{\alpha,f;a}(r)\s^a_{\alpha\beta}T^A_{ab}\psi_{\beta,f;b}(r) \right)$.  
We assume that there is single gluon exchange, where $D_{00}$ is a confining propagator,
\begin{equation}
  D_{00}({\bf p}) = \frac{\sigma_0}{(|{\bf p}|^2)^2} \; ,
\end{equation}
and $D_{\perp}$ is perturbative,
\begin{equation}
  D_{\perp}(p) = \frac{g^2}{p^2} \; .
\end{equation}
Near the Fermi surface antiquarks can be ignored, with 
$\psi^+_{\alpha,f,n}$ and $\psi_{\alpha,f,n}$ being creation and annihilation operators of
{\it nonrelativistic} fermions carrying spin (not Dirac spinors)
about the Fermi surface. Here 
$\alpha= \pm 1/2$ is a spin projection, $a,b \ldots =1,2,3$ color indices, and $f =1,...N_f$ the flavor index.
($N_f = 2$ for up and down quarks, $N_f = 3$ if we include the strange quark)
$q_f, m_f$ are electric charges and masses of the quarks,
and ${\bf A}$ is an external vector potential for QED.
The operators $T^A$ are generators of color $SU(3)$ group, $\s^a$ are Pauli matrices. 

In the first approximation we
neglect the asymmetry introduced by differences in $q_f$ and $m_f$ and by the magnetic interaction $V_2$. Then we can 
treat $j = (\alpha,f)$ as a united set of indices, so the theory (\ref{H}) is invariant under a larger symmetry of
$SU(2 N_f)$. Then realistic values are $2N_f =4$ (low density, no strange quarks) or $2N_f =6$ for high density.

If the chiral symmetry is broken, $m_f$ are renormalized  quark masses.
In fact, exact definitions of $m_f$ do not affect the qualitative side of our arguments since
for excitations near the Fermi surface the only difference between massless and massive quarks is the
difference in the Fermi velocity, $v_F$, which follows from the relationship between energy and momentum.
We also neglect the frequency dependence of the gluon propagator $D_{00}$, and set $c=1$.

The extended symmetry of $SU(2 N_f)$, instead of $SU(N_f)$, is due to a doubling from the spin
degrees of freedom, and is no longer respected once magnetic
interactions (\ref{V2}) are included.  The same symmetry has been discussed,
both in the hadronic spectrum and at nonzero quark density,
in Refs. \cite{Wagenbrunn:2007ie,Glozman:2007tv,Glozman:2015qzf,Glozman:2017vql}.

As was demonstrated previously~\cite{Kojo:2010fe}
  the quarkyonic phase supports a collection of spin-flavor density waves.  
  Since in the nonrelativistic limit spin and flavor are treated on equal footing,
  a wave with wave vector ${\bf Q}$ is characterized by  the  slow  order parameter field 
  in the form of a $2N_f\times 2N_f$ matrix field $U_{Q}$ such that  at long distances the spin-flavor density can be decomposed as
 \bea
 \rho_{jk} \equiv \sum_{n =1}^{N_c}\psi^+_{j,n}({\bf r})\psi_{k,n}({\bf r})
 - \rho_0\delta_{jk} = \sum_{{\bf Q}}U^{jk}_{Q}({\bf r})\re^{\ri {\bf Q r}}, \label{OP}
 \label{quarkyonic_condensate}
 \eea
 where $\rho_0$ is the average density. The set of wave vectors ${\bf Q}_i$
 and amplitudes of the matrix fields related to $\det U$ are determined by the matter density, which
 follows from the value of the chemical potential.
 In condensed matter systems a density wave with wave vector ${\bf Q}$ usually
 forms when the  parts of the Fermi surface connected by ${\bf Q}$ can be superimposed on each other,
 which is the nesting condition.
 Once this condition is fulfilled, the susceptibility acquires a singularity at ${\bf Q}$ signifying a possible instability.
 However, for cold quarks a perfect nesting is unnecessary
 due to the singular character of the confining potential in Eq.~(\ref{V1}) \cite{Kojo:2010fe}.
 As a result quarks can scatter with one another only at small angles and still remain near the edge of the Fermi
 surface. So it is sufficient to fulfill the nesting condition just on
 limited patches of the Fermi surface whose area, $\Lambda^2$, is determined by the interplay between
 the string tension $\s$ and the curvature of the Fermi surface.
 Within these patches the problem is essentially one dimensional,
 and can be treated by non-Abelian bosonization and conformal embedding
 \cite{DiFrancesco:1997nk, Azaria:2016}.

 In the first approximation when we neglect the magnetic interaction (\ref{V2})
 and maintain the $SU(2N_f)$ symmetry of the Hamiltonian the conformal
 embedding works as follows.  In non-Abelian bosonization the noninteracting
 one dimensional Hamiltonian can be  written as a sum of
 Wess-Zumino-Novikov-Witten terms of $U(1)$ for charge,
 $SU_3(2N_f)$ for spin and flavor, and $SU_{2 N_f}(3)$ for color
\cite{Kojo:2009ha,Kojo:2010fe,Kojo:2011cn,DiFrancesco:1997nk}.
The decomposition is adjusted to the symmetry of the interaction (\ref{V1}) which 
is given by the product of the   $SU_{2N_f}(3)$ Kac-Moody currents which commute with the first two WZNW
Hamiltonians. As a result only
the color sector experiences confinement and the other two remain massless.
They represent Abelian and non-Abelian Goldstone modes, and so the corresponding correlators 
have power law fall off at large distances.
 
We show that due to the arbitrariness of the choice of direction of the ${\bf Q}$'s,
the modes which have a linear spectrum in  one dimension acquire a
quadratic dispersion in the direction along the Fermi surface  when transverse derivatives are included.
This coupling together with the magnetic interaction (\ref{V2})
also breaks the   extended symmetry from $SU(2 N_f)$ down to $SU(N_f)$.  

The minimal possible number of patches is six, as a
cube embedded into a spherical Fermi surface.
When the density increases it becomes energetically advantageous to form triangular patches at the corners of the cube,
so another eight patches, with fourteen  in all.  
 
We estimate the number of patches as follows.
The effective current-current interaction in (\ref{H}) scales to strong coupling
giving rise to a characteristic energy scale, $\Delta$.
This can be estimated from the self consistency condition
in the rainbow diagram with one gluon and one quark propagator:
\bea
\Delta({\bf Q}) &=& \int \frac{\rd\omega \rd^3q}{(2\pi)^4}D_{00}({\bf q})
\la\la\psi(\omega,{\bf Q}/2 + {\bf q})\psi^+(\omega, -{\bf Q}/2 + {\bf q})\ra\ra  \\
&\sim &\s_0 \int_{\Delta}\frac{\rd^2 q_{\perp}\rd q_{\parallel}}{(q_{\perp}^2 + q_{\parallel}^2)^2}
\int \frac{\rd\omega \Delta}{\omega^2 +(v_Fq_{\parallel})^2} \sim \frac{\s_0 v_F}{\Delta} \; ,
\eea
where $v_F$ is the Fermi velocity and $|{\bf Q}| = 2k_F$.
The size of the patch is estimated setting the transverse part of the quarks' kinetic energy equal to this scale:
 \be
 p_{\perp}^2/2m = \Delta, ~~ \Lambda^2 = \pi p^2_{\perp} \sim m\sqrt{v_F\s_0}. \label{size}
 \ee
 From (\ref{size}) we find  the following estimate for the number of patches:  
 \be
 N_{patches} \sim \frac{k_F^2}{\Lambda^2} \sim \sqrt{\frac{n}{m\s_0}}, \label{number}
 \ee
 where $n$ is the density of quarks. Since $\sigma_0$ is essentially zero above some
 value of the chemical potential, for massive quarks
 the number of patches first grows and than sharply decreases with density.
 The estimate of Eq.~(\ref{number}) also shows the dependence of the number of patches on the quark mass $m$.
 For massless quarks, $p_{\perp} \sim \Delta$, and $\Lambda^2 \sim \sigma_0$.  Thus chiral symmetry
 breaking decreases the number of patches.
 
 The quantum action for an individual patch describes the baryonic strange metal described in Ref. \cite{Azaria:2016}:
 the quarks are confined, but the baryons remain gapless and incoherent.
 The only coherent excitations are the bosonic collective modes, so that the corresponding phase can be characterized
 as a Bose metal.

\section{Ginzburg- Landau description of Quarkionic Crystal Phase}
\label{ginzburg_landau}

In this Section  we  return to the problem of description of the Quarkyonic Crystal phase.
We show that at nonzero temperature there is no long range order and
that this state resembles smectic  liquid crystals,
which are ordered periodically only in one direction, although  with some significant caveats. 
 
As we have mentioned above, the action for the fluctuations normal to the Fermi surface is given
by the sum of WZNW actions for each patch. For the static components of the order parameter fields
one can omit the Wess-Zumino terms, leaving just terms from the gradient expansion.
The corresponding  Ginzburg-Landau (GL) free energy  for the fields $U$ was written in \cite{Kojo:2010fe},
but it requires correction. It is known
\cite{Kleinert:1981zv,Baym:1982ca,Kolehmainen:1982jn,Bunatian:1983dq,Takatsuka:1987tn,Kleinert:1999gc,Lee:2015bva,Hidaka:2015xza,Nitta:2017mgk,Nitta:2017yuf}
that the fluctuations tangential to the Fermi surface must have a zero stiffness,
since the orientation of the entire set of ${\bf Q}_i$'s is arbitrary and for spherical Fermi
surface any such rotation costs zero energy.
Therefore at nonzero temperature where all fluctuations are classical, the
free energy density similar to that in smectic liquid crystal  \cite{Lee:2015bva}:
 \bea
 {\cal F}/T &=& \frac{1}{2T}\sum_{{\bf q}}\Big\{ \tilde\lambda_{1,Q} \; 
 \mbox{Tr}({\bf q}\vec\nabla U_Q)({\bf q}\vec\nabla U_Q^+) \nonumber\\
 &+ &  \tilde\lambda_{2,Q}\; \mbox{Tr}[({\bf q}\times\vec\nabla)^2 U_Q)][({\bf q}\times\vec\nabla)^2 U^+_Q]\Big\}
 + {\cal V}(U^+,U) , ~~ {\bf q} = {\bf Q}/Q \;, 
 \label{action1}
 \eea
 where $U_{-{\bf Q}} = U^+_{{\bf Q}}$ and ${\cal V}$ is the local potential which
 fixes the amplitudes of these matrix fields.
 We normalize the ${\bf Q}$'s as unit vectors. Eq.~(\ref{action1}) can be formally derived from Eq. (\ref{H}).
 The first term was derived in our  previous paper \cite{Kojo:2010fe}, while
 the second originates from the fusion of the two perturbing operators
 \bea
\hat T_{\perp} &=& - \frac{v_F}{2k_F}(R^+\nabla_{\perp}^2R + L^+\nabla_{\perp}^2L), \\
R({\bf p}) &=& \psi({\bf Q}/2 + {\bf p}), ~~ L({\bf p}) = \psi(-{\bf Q}/2 + {\bf p}) \; . 
\eea
Then the  estimates of the parameters: 
$ \lambda_1 \sim  v_F\Lambda^2, ~~ \lambda_2/\lambda_1 = Ck_F^{-2}, $ 
 where $C$ is a numerical constant and $k_F$ is the Fermi wave vector. 

 When the Fermi vectors of up, down and strange quarks are different, the GL theory should be augmented by the term
 \bea
 && (m_s - m_{u,d})c^2\int \rd^3 x \;
 \psi^+_{\s,f,n}\psi_{\s,f,n} =  \ri (m_s - m_{u,d})c^2 \int \rd^3 x \frac{N_f}{\pi}
 \mbox{Tr}\Big[\hat\tau^3 U_{Q}({\bf q}\vec \nabla)U_{\bf Q}^+\Big] \; , \nonumber\\
 &&  \tau^3 = \mbox{diag}(-1,-1,2)\otimes \hat I,
 \eea
 where the first matrix in the tensor product acts in flavor space, and the second in spin space.  This contribution
 is the non-Abelian bosonization of the above fermionic term.
 This extra contribution can be removed by the redefinition of the matrix field:
 \be
 U_{Q} \rightarrow \re^{\ri \delta\mu \tau^3 (\vec Q\vec r)}U_{Q}.
 \ee
 So the shift of the Fermi level of strange quarks does not break the SU(6) symmetry, at least
 in the leading approximation in $\delta m$. A magnetic field does, which is taken into account later.
 
 The matrix $U$ can be parametrized  as
 \be
 U_Q = A_Q\re^{\ri\phi_Q}G_Q, \label{A}
 \ee
 where the amplitude $A_Q$ is fixed by the potential ${\cal V}$ (\ref{action1}),
 and $G_Q$ is a $SU(2N_f)$ matrix. Omitting the massive fluctuations of the amplitude $A_Q$ we get from (\ref{action1}) 
 the free energy density for the soft modes. It is divided into two parts, Abelian and non-Abelian sigma models:
 \bea
 && \frac{\cal F}{T} =S_{U(1)} + S_{SU(2N_f)}, \\
 && S_{U(1)}= \frac{1}{2T}\sum_{{\bf Q}}\Big[\lambda_1({\bf q}\cdot \vec\nabla \phi_Q)^2
 + \lambda_2[({\bf q}\times\vec\nabla)^2 \phi_Q)]^2\Big],\label{Gauss}\\
 && S_{SU(2N_f)}= \frac{1}{2T}\sum_{{\bf Q}}\Big\{ \lambda_1
 \mbox{Tr}({\bf q} \cdot \vec\nabla G_Q)({\bf q}\cdot \vec\nabla G_Q^+) +
 \lambda_2\mbox{Tr}[({\bf q}\times\vec\nabla)^2 G_Q)][({\bf q}\times\vec\nabla)^2 G^+_Q]\Big\}.  \label{SU(2)}
 \eea
 %

\section{ Fluctuations and order} 
 \label{fluctuations}
 
 We next show that because the transverse stiffness vanishes, no symmetry is broken at nonzero temperature.
 In that respect the Quarkyonic Phase resembles the lamellar phases  of liquid crystals,
 which have the same bare fluctuation spectrum. 

 Under renormalization, the Abelian, Eq.~(\ref{Gauss}), and non-Abelian, Eq.~(\ref{SU(2)}),
 parts of the free energy behave very differently.
 The Abelian part is just a free theory,
 since the effects of vortices in three dimensions can be ignored.
 On the other hand, due to the softness of the transverse fluctuations,
 the action of Eq.~(\ref{SU(2)}) renormalizes to strong coupling, which generates a finite correlation length.
 In the one loop approximation the renormalization group equations do not differ from those for a 
 two-dimensional $SU(2N_f$)-symmetric  Principal Chiral Field model.
 The analogy becomes clearer when one uses the saddle point approximation.  
 A rough  estimate for the correlation length  can be obtained if we replace the
 local constraint for the matrix field $GG^+ = I$
 by its average $\la GG^+\ra =1$.
 We enforce the constraint $GG^+ = I$ by adding to the action the term $i \eta(G^+G -I)$,
 where $\eta$ is the Lagrange multiplier field.
 In the  saddle point approximation we replace the multiplier field  $\eta$
 by a constant $i \eta = \lambda_1\xi^{-2}$.
 Then we extract  the propagator of $G$ from Eq.~(\ref{SU(2)}), as
 the constraint $\la GG^+\ra =1$ yields the equation for the inverse correlation length $\xi^{-1}$:
 \bea
 1 = \frac{4N_fT}{(2\pi)^3}\int
 \frac{\rd k_{\parallel}\rd^2 k_{\perp}}{\lambda_1k_{\parallel}^2 +\lambda_2k^4_{\perp} + \lambda_1\xi^{-2}} \; ,
 \eea
 where the correlation length $\xi$ is 
 \be
 \xi \sim v_F\Lambda^{-1}\exp\Big( \pi\sqrt{\lambda_1\lambda_2}/ N_fT\Big) = v_F\Lambda^{-1}\exp\Big(C\frac{\Lambda^2}{m N_fT}\Big), 
\ee
where $C \sim 1$.
Thus the non-Abelian sector is disordered, although at low temperature, the correlation length is exponentially
large.  The Abelian action in Eq.~(\ref{Gauss}) is a free theory.
The corresponding observables are complex exponents of $\phi_Q$ and as such are periodic functionals of $\phi_Q$.
In two spatial dimensions vortices of the $\phi_Q$ field would also enter, but
in three spatial dimensions these vortices are extended objects with an energy proportional to their length,
and so can be ignored.  Thus at nonzero temperature, there is a phase transition of second order
into a phase characterized by long range correlations of the fields $\phi_Q$.

An order parameter can be constructed for this critical phase.  It cannot directly involve
correlations of $\rho_{jk}$, because the correlations of such fields decay exponentially.
However, since the charge phase $\phi_Q$ is a free field over large distances, we can construct
an operator which exhibits quasi long range order.
Since $\det G =1$ the order parameter includes $4N_f$ fermions:
  \be
  {\cal O}_{\bf Q} = \det\hat\rho_{\bf Q} = \re^{2\ri N_f({\bf Qr}+ \phi_Q)}B_{{\bf Q}}.
  \label{quarkyonic_order}
 \ee
 At distances $\gg \xi$ when fluctuations of $G$ can be treated as massive one
 can replace $B$ by some  fixed amplitude. The average $\la {\cal O}_Q\ra =0$,
 but its correlations fall off as powers of the relative distance:
\bea
&& \la\la {\cal O}_{\bf Q}({\bf r}_1){\cal O}^+_{\bf Q}({\bf r}_2\ra\ra \sim 
\frac{\cos[2N_f{\bf Qr}_{12}]}{\Big\{({\bf qr}_{12})^2 + k_0^2[{\bf q}\times{\bf r}_{12}]^4\Big\}^{d}},\nonumber\\
&&  ~~ d = N_f^2Tk_0/\pi\lambda_1, ~~ k_0 \sim k_F.  
\eea
So  at finite temperatures the quarkyonic crystal melts into
an Abelian critical phase with  wave vectors $2N_f$ times greater
than the ones established by the energetics at zero temperature.
It is essentially a density wave of a quasi-condensate of $4N_f$ bound states of quarks.
These bound states are spin and flavor singlets.

\section{ Magnetic field} 
\label{magnetic.tex}

A sufficiently strong  magnetic field
\cite{Skokov:2009qp} has a profound effect on the structure of the quarkyonic phase.
We will start our analysis at zero temperature when the $SU(2N_f$) symmetry is spontaneously broken.
In this case the matrix $G_Q$ can be approximated as
 \be
 G_Q(r) = G_0 \; \re^{i \, t_Q^F(r) \, \hat{T}^F} \approx G_0(1 + i \, t_Q^F(r) \, \hat{T}^F) \; ,
 \ee
 where $G_0$ is some constant matrix and $\hat T^F$ are generators of the $SU(4)$ algebra.
 Then to leading order, instead of Eq.~(\ref{SU(2)}) to quadratic order we obtain
 \bea
 {\cal L} &=& V_0\; t^F \; \mbox{Tr}[\hat{T}^F \hat{q}(\vec\s{\bf B}) \; \hat T^{F'}
 \hat{q}(\vec\s{\bf B})]t^{F'} + \p_{\tau}\, t^{F} \, \p_{\tau} \, t^F  \nonumber\\
 &+& \lambda_1 \, t^F \, \mbox{Tr} \Big\{\hat T^F
 \Big[\lambda_1({\bf q \cdot D})^2 +\lambda_2 \, [{\bf q}\times{\bf D}]^4\Big]\hat T^{F'}\Big\} t^{F'} \; .
 \eea
$V_0 \sim \Delta^2/\mu$ is proportional to the square of amplitude of $A_Q$ in Eq.~(\ref{A}).

For $N_f =2$ it is convenient to represent the generators $T^a$ in terms of the
Pauli matrices acting in the spin and the flavor spaces:
\be
T^{s} = (\s^a\otimes I)\; , ~~T^{f} =(I\otimes\tau^a)\; , ~~T^{(s,f)} = (\s^a\otimes\tau^b) \; .
\ee
The magnetic field splits
the dispersion of the Goldstone modes. In particular, the mode  $(I\otimes\tau^z)$  is not affected by the magnetic field and remains gapless. The
three $(\s^a\otimes I)$ modes and three $(\tau^z\otimes\s^a)$ are affected only by the Zeeman term. Their spectrum is
\be
E^2 = \lambda_1 \left( p_{\parallel}^2 + \frac{p^4_{\perp}}{k_0^2} \right) + V_0 \, \sum_f q_f^2 \, {\bf B}^2 \; .
\ee
The only modes affected by the orbital magnetic field are the eight modes  with
off-diagonal $\tau^{\pm}$ Pauli matrices. To simplify the discussion of their spectrum  we will consider two limiting cases.

1. {\bf B} $\parallel$ {\bf Q}$\parallel \hat z$, $A_x = By, A_y= A_z =0$.
Then the spectrum is determined by the equation
\bea
E^2 {\bf t} = \lambda_1\Big\{p_z^2 + \frac{1}{k_0^2}\Big[ \p_y^4 + (\pm p_x -By)^4\Big]\Big\} {\bf t}.\label{ee}
\eea
Here the real vector ${\bf t}$ includes the modes corresponding to generators
$(I\otimes\tau^{\pm}), (\s^a\otimes\tau^{\pm})$.
We need just to shift $y$ by $\pm p_x/B$ after which $p_x$ drops out of the
eigenvalue equation. The general solution of Eq.~(\ref{ee}) is given by 
\bea
&& {\bf t} = \Re e\Big[{\bf f}(yB^{1/2} \pm p_x/B^{1/2})\re^{\ri(p_x x + p_z z)}\Big],
~~ E^2 = \lambda_1\Big(p_z^2 + \frac{B^2\epsilon^2}{k_0^2}\Big), \nonumber\\
&& \epsilon^2{\bf f}(\tau) = (\p^4_{\tau} + \tau^4){\bf f}(\tau).
\eea
The spectrum is 
\bea
E^2 = \lambda_1\Big[p_z^2 +\frac{B^{2}\, g_1(n)}{k_0^2}\Big] \; .
\eea
We can determine function $g_1(n)$ at large quantization numbers  $n\gg 1$ using semiclassical approximation:
\bea
{\bf f} \sim {\bf f}_0\exp\Big[\ri\int^{\tau}_{-\sqrt\epsilon}\rd t(\epsilon^2 - t^4)^{1/4}\Big].
\eea
Then the spectrum follows from the Bohr-Sommerfeld quantization condition:
\bea
\int^{\sqrt\epsilon}_{-\sqrt\epsilon}\rd t(\epsilon^2 - t^4)^{1/4} =\pi n,
\eea
so that at $n\gg 1$ 
\bea
g_1(n) = (\gamma n)^2,  ~~ \gamma = \pi \left/\int_{-1}^1(1- x^4)^{1/2}\rd x \right. \; .
 \eea

 2. {\bf B} $\perp$ {\bf Q} and $\parallel \hat y$, $A_x = Bz, A_y=A_x =0$.
\bea
E^2 {\bf t} = \lambda_1\Big\{- \p_z^2 + k_0^{-2}\Big[(p_x \pm Bz)^4 + p_y^4\Big]\Big\} {\bf t}.
\eea
The spectrum is given by 
\bea
 E^2 = \lambda_1k_0^{-2}\Big[p_y^4 + (k_0B)^{4/3}g_2(n)\Big], 
 \eea
 where using the semiclassical approximation at $n\gg 1$ we get  
 \bea
 g_2(n) = (\eta n)^{4/3}, ~~ \eta = \pi\left/\int_{-1}^1(1- x^4)^{1/2}\rd x \right. \; .
\eea

To summarize: in the presence of a magnetic field, the spectrum is divided into three groups. There are eight gapped  modes which include
components along the $\tau^{\pm}$ generators; their gaps depend strongly
on the direction of the magnetic field.  The second group consists of the  modes which
include only spin operators and $\tau^z$; they  have much smaller  gaps $\sim B(\Delta/\mu)$.
The third group includes the mode $(I\otimes\tau^z)$ which remains gapless.

Hence there are three regimes of temperature. When temperature is so  high
that the correlation length is smaller than the magnetic length the influence of the magnetic field is small:
\be
\exp(- C\Lambda^2/N_f mT) > B\mu_B/k_0v_F. 
\ee

There is an  intermediate interval of temperature
when the magnetic field suppresses some modes, leaving as  nominally gapless the modes
$(I\otimes\tau^z), (\s^a\otimes I)$ and $(\s^a\otimes\tau^z)$.
In this region one can approximate the order parameter field as a product of
SU(2) matrix $g$ and U(1) matrix V:  $G = g_{SU(2)}V$ with $V = \cos\alpha I+ \ri \tau^z\sin\alpha$.
The SU(2)-symmetric part of the order parameter is disordered by thermal fluctuations,
as was demonstrated in Sec. (\ref{fluctuations}).
The action for the U(1) part is Gaussian and the field $\alpha$ has long range correlations.
This is in addition to the overall U(1) phase $\phi$.

At yet smaller temperatures the SU(2) part $g$ is gapped by the magnetic field, through quantum effects.
In both cases the  Abelian modes $\alpha,\phi$ (the total charge and the diagonal flavor one) survive as gapless.
Being Abelian they remain long ranged even if fluctuations are taken into account.
As a consequence by gapping out all non-Abelian modes the magnetic field
instigates quasi long range order in the $\sum_{\s} \rho_{(f,\s),(f,\s)}$
which is quadratic in quark creation and annihilation operators.

\section{Phase diagrams with a Lifshitz point}
\label{phase_diagram}
\subsection{General effective Lagrangian}
\label{general_phase}

In the previous section we considered
chiral spirals in a quarkyonic phase in QCD.  This is relevant at chemical potentials
above that for hadronic nuclear matter, but below those where perturbative QCD applies.

Moving up in chemical potential,
the transition from a quarkyonic phase to the perturbative regime
appears straightforward.  The width of a patch with a chiral spiral is proportional to the square root of the string
tension, Eq.~(\ref{size}).  As discussed in Sec. (\ref{model_quarkyonic}), 
numerical simulations for two colors show that the string tension decreases with increasing
chemical potential, and is essentially zero by
$\mu_q \sim 750$~MeV, Fig.~3 of Ref. \cite{Bornyakov:2017txe}.
In a perturbative regime the interactions near the Fermi surface lead to color superconductivity
symmetrically over the Fermi surface, instead of leading to formation of chiral
spirals in patches.  On the quarkyonic side the only order parameter is the $U(1)$ phase of the determinantal
operator, $\det\hat\rho_{\bf Q}$ of Eq.~(\ref{quarkyonic_order}).
On the color superconductivity side the broken symmetry is completely different and therefore 
it is natural
to expect that the phase transition is of first order,
as between charge density wave and superconductivity in condensed matter systems.
There are also other order parameters for color superconductivity which may enter \cite{Pisarski:1999gq}.
For three colors the precise value at which this transition occurs for three colors cannot be fixed by our qualitative
arguments, but following the discussion at the beginning of Sec. (\ref{model_quarkyonic}), it is
presumably somewhere around $\mu_q \sim 1$~GeV.

Going down in density, from the quarkyonic phase to hadronic nuclear matter, there are chiral
spirals in the former, and pion/kaon condensates in the latter.  In terms of the usual chiral order
parameter, 
\begin{equation}
  \rho= \overline{q}_L \, q_R \; ,
  \label{define_chiral}
\end{equation}
the condensate between $\sigma$ and $\pi^3$ is given by
\begin{equation}
\rho_{\pi \, {\rm cond.}} = \rho_0 \; \exp\left( i \, (\bold{Q} \cdot \bold{r} + \phi) \, t_3 \right) \; ,
\label{pion_condensate}
\end{equation}
where $\rho_0$ is a constant $\sim f_\pi$,
and $t_3$ is a flavor matrix.  In neutron stars, with a charged background of protons the
analogous condensates are along the directions corresponding to $\pi^-$ and $K^-$
\cite{Overhauser:1960,Migdal:1971,Sawyer:1972cq,Scalapino:1972fu,Sawyer:1973fv,Migdal:1973zm,Migdal:1978az,Migdal:1990vm,Kleinert:1981zv,Baym:1982ca,Kolehmainen:1982jn,Bunatian:1983dq,Takatsuka:1987tn,Kleinert:1999gc,Kaplan:1986yq,Brown:1993yv,Brown:1995ta,Brown:2007ara}.

This suggests that there is a direct relation between the pion/kaon condensates of hadronic
nuclear matter and those of the quarkyonic regime.  We argued previously that the only true order
parameter for the quarkyonic regime was an overall phase factor of $U(1)$.  Such a phase clearly arises
for the field of Eq.~(\ref{pion_condensate}), which we denote by the phase $\phi$.  The physical
origin of this phase is obvious: at a given point along the $\bold{Q}$ direction, the condensate
points entirely in a given direction, say along $\pi^3$.  Where this point is just an overall shift in the phase,
though.

Thus pion/kaon condensates have a $U(1)$ phase, which is sometimes termed the ``phonon'' mode
\cite{Tatsumi:2014cea,Hidaka:2015xza}.  This is then a strict order parameter which
distinguished hadronic nuclear matter, without a pion/kaon condensate, from that with.
If this transition is not of first order, then it must be of second order, in the universality
class of $U(1)$.

As the chemical potential increases further, it is very plausible that it is not possible to rigorously
distinguish between the pion/kaon condensate of Eq.~(\ref{pion_condensate}) and
chiral spirals of the quarkyonic regime.  The only difference is
that there are $N_f^2 -1$ very light modes for a pion/kaon condensates, and $4 N_f^2 - 1$ light
modes for a quarkyonic chiral spiral.  It is natural to assume that the additional $3 N_f^2$ modes of the latter become lighter as $\mu_q$
increases.  

As discussed previously, there may be phase transitions as the number of patches increases, although
this is not really necessary.  In that vein, we note that in the original discussion of pion condensates
by Overhauser \cite{Overhauser:1960}, it was explicitly stated that the simplest solution in three
dimensions is that with six patches.

The relation between pion/kaon condensates and chiral spirals also suggests a less trivial
speculation.  For static quantities, at high density the effective theory 
for the light modes of a chiral spiral is a $SU(2 N_f)$ sigma model.  Once transverse fluctuations
(for massless quarks) are included, or magnetic interactions (for massive), as we argued in Sec.
(\ref{ginzburg_landau}), $SU(2 N_f)$ sigma model reduces to a $SU(N_f)$ model.
This agrees with the effective theory for a pion/kaon condensate, which is a nonlinear sigma
model on $SU(N_f)$.

However, we argued that for non-static quantities, there is also a WZNW term, of level
$3$ for $SU(2 N_f)$, and level $6$ for $SU(N_f)$.  This suggests that there might be
a WZNW term for pion/kaon condensates, with level $6$.  This is not obvious.  The original
theory has a WZNW term in four dimensions, but this involves derivatives in all four dimensions \cite{Witten:1983tw}.
Perhaps a WZNW term arises in two dimensions from the variation of the patches
in the transverse directions.

The similarity between chiral spirals
and pion/kaon condensates has been recognized previously,
at least implicitly \cite{Buballa:2014tba,Heinz:2013hza,Heinz:2015lua,Lee:2015bva}.
The appearance of a $U(1)$ order parameter, and the possible appearance of a WZNW term, is novel.

To support the above considerations we establish a formal correspondence between the relativistic and the
nonrelativistic versions of the QCD Hamiltonian. 
We start with the Dirac Hamiltonian:
\begin{equation}
H = \hat q^+({\bf p})\Big(\hat\tau_z\otimes\hat\s^a p^a + m\hat\tau^x\otimes\hat I -\mu \hat I\otimes\hat I\Big)\hat q({\bf p}), \label{Dirac}
\end{equation}
where $\tau^a$ act in the the chiral basis $(R,L)$ and $\s^a$ act in the spin space. We neglect the quark mass. 
Then under the transformation
\bea
\left(
\begin{array}{c}
 q_{R\s}({\bf p})\\
q_{L\s}({\bf p})
\end{array}
\right) &=& \frac{1}{\sqrt 2}\left(
\begin{array}{c}
z_{\s}({\bf n})\\
\epsilon_{\s\s'}z^*_{\s'}({\bf n})
\end{array}
\right)\psi_+(p) + \frac{1}{\sqrt 2}\left(
\begin{array}{c}
\epsilon_{\s\s'}z^*_{\s'}({\bf n})\\
z_{\s}({\bf n})
\end{array}
\right)\psi_-({\bf p})
   \nonumber\\
  &+& \frac{1}{\sqrt 2}\left(
\begin{array}{c}
z_{\s}({\bf n})\\
-\epsilon_{\s\s'}z^*_{\s'}({\bf n})
\end{array}
\right)\eta^+_-(p) + \frac{1}{\sqrt 2}\left(\begin{array}{c}
\epsilon_{\s\s'}z^*_{\s'}({\bf n})\\
- z_{\s}({\bf n})
\end{array}
\right)\eta^+_+({\bf p}) \; ,\label{dirac}
\eea
where
\be
z^+\vec\s z = {\bf n} \equiv \frac{{\bf p}}{p}, ~~z_{\s}({\bf n}) =
\epsilon_{\s\s'} z^*_{\s'}(-{\bf n}), ~~ \sum_{\s = \pm 1}z^*_{\s}z_{\s} =1. 
\ee
the Hamiltonian Eq.~(\ref{Dirac}) becomes
\bea
 H = \sum_{{\bf p},\tau =\pm} \Big[(|p|-\mu)\psi_{\tau}^+({\bf p})\psi_{\tau}({\bf p}) +(|p|+\mu)\eta_{\tau}^+({\bf p})\eta_{\tau}({\bf p})\Big].
 \eea
In what follows we will drop the antiparticles $\eta$. 

Now let us consider the chiral order parameter $\rho$ selecting in it only the part associated with particles: 
\bea
\rho =  q^+(r)\tau^x q &=& \frac{1}{2}\sum_{p,p'}\re^{\ri({\bf p}-{\bf p}'){\bf r}}
\psi^+_{\alpha}(p)\psi_{\beta}(p') \Big({\bf e}^*_{\alpha}(p){\bf e}_{\beta}(p')\Big) + ...  \nonumber\\
 &=& \frac{1}{2}\sum_{Q,p}\re^{\ri {\bf Qr}}\psi^+_{\alpha}({\bf Q}/2 +{\bf p})\psi_{\alpha}(-{\bf Q}/2 +p) + ..., \label{osc}
 \eea
 where the basis vectors ${\bf e}$ are defined in (\ref{dirac}) and the ellipses 
 include the contributions of antiparticles.
 This shows that the chiral order parameter $\rho$, which is {\it uniform} at
 $\mu =0$, naturally acquires oscillatory terms at $\mu \neq 0$.
   These can be either pion/kaon condensates or
 quarkyonic chiral spirals.  As we neglected antiparticles, this only happens for sufficiently large $\mu_q$.
 These arguments do not show how large $\mu_q$ must be, but they do establish
 that both chiral symmetry breaking and the formation of
 inhomogenous phases can be described within the same effective model.

To discuss the phase diagram we then consider a $SU(N_f) \times SU(N_f)$ field $\rho$,
taking a customary linear sigma model,
$$
{\cal L} = \frac{1}{2} \; {\rm tr} \, |\partial_0 \, \rho |^2
+ \; \frac{\sqz}{2} \; {\rm tr} \, |\partial_i \, \rho |^2
+ \; \frac{1}{2 \, {\cal M}^{2}} \; {\rm tr} \, |\partial^{2} \rho |^2
$$
\begin{equation}
  + \; \frac{m^2}{2} \; {\rm tr} \, \rho^\dagger \rho
  + \; \frac{\lambda_1}{4} \; \left({\rm tr} \, \rho^\dagger \rho \right)^2
  + \; \frac{\lambda_2}{4} \; {\rm tr} \, (\rho^\dagger \rho)^2
  + \; \frac{\kappa}{6} \; ({\rm tr} \, \rho^\dagger \rho)^3 + \ldots 
  \; .
  \label{eff_lag}
\end{equation}
In four spacetime dimensions $\rho$ has dimensions of mass.
The first two terms are standard kinetic terms.
The coefficient of the second term, with two spatial derivatives, can
have an arbitrary coefficient $\sqz$.
Implicitly we consider systems at nonzero temperature and quark density, and so
there  is a preferred rest frame.  Then Lorentz symmetry is lost, and $\sqz \neq 1$ is allowed.
In particular, we allow $\sqz$ to be negative.   Stability then requires the addition of a positive
term with four spatial derivatives; the coefficient of that term is
$\sim 1/{\cal M}^2$, where $\cal M$ is some mass scale derived from the underlying theory.

While we only consider static quantities, and so can ignore the first term with two time
derivatives, we add it 
to emphasize that the {\it only} higher derivatives considered are those in the spatial coordinates.
It is well known that higher order derivatives in time lead to acausal behavior, which should
not occur in an effective theory.
This is not dissimilar to  causal theories of higher derivative gravity, such
as Horava-Lifshitz gravity \cite{Horava:2009if,Horava:2009uw,Mukohyama:2010xz,Sebastiani:2016ras}.

When the coefficient of the term with two spatial derivatives is positive,
$\sqz > 0$, the phase diagram is standard.  For positive mass squared,
$m^2 > 0$, the theory is in a symmetric phase, with $\langle \rho \rangle = 0$.  For positive
quartic coupling, $\lambda > 0$, the global flavor symmetry is broken when 
the mass squared is negative, $m^2 < 0$, but not the translational symmetry.
At zero mass, $m^2 = 0$, there is a second order transition in the appropriate universality class.

The hexatic couplings, such as $\kappa$, are assumed to be positive, so the
quartic coupling $\lambda$ can be negative.  It is easy to show
that there is then a first order transition at positive mass squared, $m^2 > 0$.

In the plane of the mass squared, $m^2$, and the quartic couplings $\lambda$, the phase diagram
is standard.  For positive $\lambda$ there is a line of second
order transitions when $m^2 = 0$.  For negative $\lambda$ there is a line of first
order transition at positive $m^2$.  These two lines meet at the origin, $m^2 = \lambda = 0$,
which is a tricritical point.  
In three dimensions the critical exponents for a tricritical point are those of mean field,
with logarithmic corrections controlled by the hexatic interactions \cite{Amit:2005}.

Since there is more than one quartic coupling, 
even if the quartic couplings are originally positive, in the infrared limit
they can flow to negative values, and so generate a fluctuation induced first order transition 
\cite{Coleman:1973jx,Halperin:1973jh,Bak:1976zza,Amit:2005}.
For this Lagrangian in $4 - \epsilon$ dimensions, to leading order in $\epsilon$
this happens when $N_f > \sqrt{2}$ \cite{Pisarski:1983ms}.
This is only valid to leading order in $\epsilon$.  For two flavors, 
the symmetry is $SU(2)_L \times SU(2)_R \times U(1)_A \equiv O(4) \times O(2)$, then 
the conformal bootstrap program suggests that there is a non-trivial fixed point
in three dimensions, $\epsilon = 1$, which is not present for small $\epsilon$
\cite{Nakayama:2014lva,Nakayama:2014sba}.  If true, then in the plane of $m^2$ and $\lambda$,
there are two lines of first order transitions which meet at a tricritical point.

We next consider the corresponding phase diagram in the plane of $m^2$ and the wave
function renormalization constant $\sqz$.  The case of positive $\sqz$ is  a
trivial consequence of the above analysis, with a line of second order transitions along
$m^2 = 0$.

We next turn to the case of $m^2 < 0$, when $\rho$ has an expectation value $\rho_0 \neq 0$.
Spatially inhomogenous condensates arise when $\sqz$ is negative.
We take
\begin{equation}
  \rho_{CS} = \rho_0 \; \exp( i {\bold Q} \cdot {\bold x} ) \; ,
  \label{two_comp_cond}
\end{equation}
where ${\bold Q} = Q_0 \hat{z}$.

The ansatz of Eq.~(\ref{two_comp_cond}) is only a caricature of the full solution.
The detailed form of the condensate differs
depending upon whether the broken symmetry is discrete or continuous.
For a discrete $Z(2)$ symmetry, the condensate is a kink crystal, where
the field oscillates in sign in one direction,
as in Eq.~(\ref{two_comp_cond}).
For a continuous symmetry of $U(1)$, the condensate is a 
spiral, where the field oscillates in two directions, as in Eqs. (\ref{pion_condensate}).
These differences
are illustrated by the exact solutions in $1+1$ dimensions: for an infinite number of flavors,
at low $T$ and nonzero $\mu$
the Gross-Neveu model develops a kink crystal, while the chiral Gross-Neveu model has a chiral
spiral \cite{Schon:2000he,Schnetz:2004vr,Thies:2006ti,Basar:2008im,Basar:2008ki,Basar:2009fg}.
The condensates for more complicated continuous symmetries can be more involved,
as for Eq.~(\ref{quarkyonic_condensate}).  

For our qualitative analysis, though, the precise form of the condensate is secondary.
We then minimizing the terms with spatial derivatives with respect to $Q_0$,
\begin{equation}
  Q_0^2 = - \, 2 \, \sqz \, {\cal M}^2 \; .
  \label{k0_min}
\end{equation}
For $Q_0$ to be real, $\sqz$ has to be negative.  
Plugging this back into the Lagrangian, we obtain
\begin{equation}
  {\cal V}  = \frac{1}{2} \, \meff^2 \, \rho^{\, 2}
    + \; \frac{\lambda}{4} \; (\rho^{\, 2})^2 \; .
\label{potential_k0}
\end{equation}
where $\lambda = N_f^2 \lambda_1 + N_f \lambda_2$, and
\begin{equation}
  \meff^2 = m^2 - \,  \frac{\sqz^2}{4} \, {\cal M}^2 \; .
  \label{eff_mass}
\end{equation}
Henceforth we ignore the hexatic coupling $\sim \kappa$, as we uniformly assume that
the quartic coupling $\lambda$ is positive.

Consider a given value of $m^2 < 0$, crossing 
from positive to negative values of $\sqz$.  For $\sqz > 0$, the theory is
in a broken phase; for negative $\sqz$, in the phase with the chiral spiral of
Eq.~(\ref{two_comp_cond}).   Ignoring $\kappa$, the potential energy at the minimum is
\begin{equation}
  {\cal V} = - \frac{1}{4 \lambda} \, \meff^4 \; .
  \label{potential_minimum}
\end{equation}
Assuming that the variation in $\sqz$ is linear in the appropriate thermodynamic variable, such
as the chemical potential or temperature $T$,
\begin{equation}
  \sqz = z_0 \, (T - T_0) \; ,
  \label{mean_field_z}
\end{equation}
then it is trivial to show that while the potential, or free energy, is continuous at $T = T_0$, the
first derivative, related to the energy density, is discontinuous.
This is natural: except for Goldstone bosons, the correlation lengths are nonzero in both phases.
Further, there is an order parameter which distinguishes the phases: $\langle \phi \rangle$
is constant when $\sqz$ is positive, while with the chiral spiral of Eq.~(\ref{two_comp_cond}),
the spatial average of $\langle \phi \rangle$ vanishes when $\sqz < 0$.

Consider next the case when $\sqz$ is negative and the original mass squared, $m^2$, is positive.  Then
the effective mass $\meff$ vanishes when $m^2 = \sqz^2 {\cal M}^2/4$, and one expects a second
order phase transition. 

It was shown by Brazovskii \cite{Brazovski:1975,Dyugaev:1982gf,Hohenberg:1995,Karasawa:2016bok}
that instead there is a first order transition.  For the assumed parameters, the $\phi$ propagator is
\begin{equation}
  \Delta^{-1}(\vec{k}) = m^2 + \sqz \, \vec{k}^2 + \frac{(\vec{k}^2)^2}{{\cal M}^2} \; .
  \label{braz_prop}
\end{equation}
When $\sqz < 0$, there is a minimum for nonzero spatial momentum.  Expand
\begin{equation}
  \vec{k} = (k_{{\rm tr}}, Q_0 + k_z) \; .
  \label{shiftk}
\end{equation}
Inserting Eq.~(\ref{shiftk}) into Eq.~(\ref{braz_prop}) and expanding, the terms
proportional to $k_z$ vanish if $Q_0$ satisfies Eq.~(\ref{k0_min}).  Then
\begin{equation}
  \Delta^{-1}(\vec{k}) \approx m_{{\rm eff}}^2 - \, 2 \, \sqz \, k_z^2 + \ldots ,
  \label{prop_sym}
\end{equation}
where $m_{{\rm eff}}^2$ is that of Eq.~(\ref{eff_mass}).  Notice that the terms quadratic
in the transverse momenta, $k_{{\rm tr}}$, vanish, although there
are terms of higher order in $k_{{\rm tr}}$. This is similar to the behavior of
Goldstone bosons in a chiral spiral. 

Consequently, an integral over virtual fields is dominated by fluctuations
in the direction of $k_z$.  When the effective mass is small,
the correction to the mass term is
\bea
\Delta m^2 \sim \lambda T\int \frac{d^3 k}{(2\pi)^3} \;
\frac{1}{\meff^2 + (k^2- Q_0^2)^2/{\cal M}^{2} } \sim + \frac{T\lambda Q_0{\cal M}}{\meff} \; .
\label{mass_braz}
\eea
Similarly, the correction to the quartic coupling is 
\begin{equation}
  \Delta \lambda \sim - \lambda^2 T \; \int d^3 k \; \frac{1}{(\meff^2 + (k^2- Q_0^2)^2/{\cal M}^{2})^2 }
  \sim - \; \frac{\lambda^2 \, T Q_0 {\cal M}}{\meff^3} \; .
  \label{coupling_braz}
\end{equation}
Both of these results follow because the fluctuations for small $\meff$ are those for a theory in
{\it one} dimension, along $k_z$.  Because the infrared divergences of Eqs. (\ref{mass_braz}) and
(\ref{coupling_braz}) bring in powers of $1/\meff$, a second order transition, where
$\meff = 0$, is not possible.  There is a transition between the two phases,
but it is necessarily of first order, where $\meff$ is always nonzero in each phase.

This has been termed a ``fluctuation induced first order'' transition
\cite{Brazovski:1975,Dyugaev:1982gf,Hohenberg:1995,Karasawa:2016bok}, but the
terminology is somewhat misleading.  In theories with several coupling constants, 
couplings can flow to negative values 
\cite{Coleman:1973jx,Halperin:1973jh,Bak:1976zza,Amit:2005}, and so generate
a first order transition.  This depends upon how the coupling constants flow
under the renormalization group in the infrared limit, and so depends both
upon the symmetry group, and the dimensionality of space-time.  

In contrast, what happends for $m^2 > 0$ and $\sqz < 0$ is
just an effective reduction of the fluctuations to one dimension.
It does not depend upon either the global symmetry or the original dimensionality
of spacetime.  

\begin{figure}[t]
	\centerline{  \includegraphics[scale=0.5]{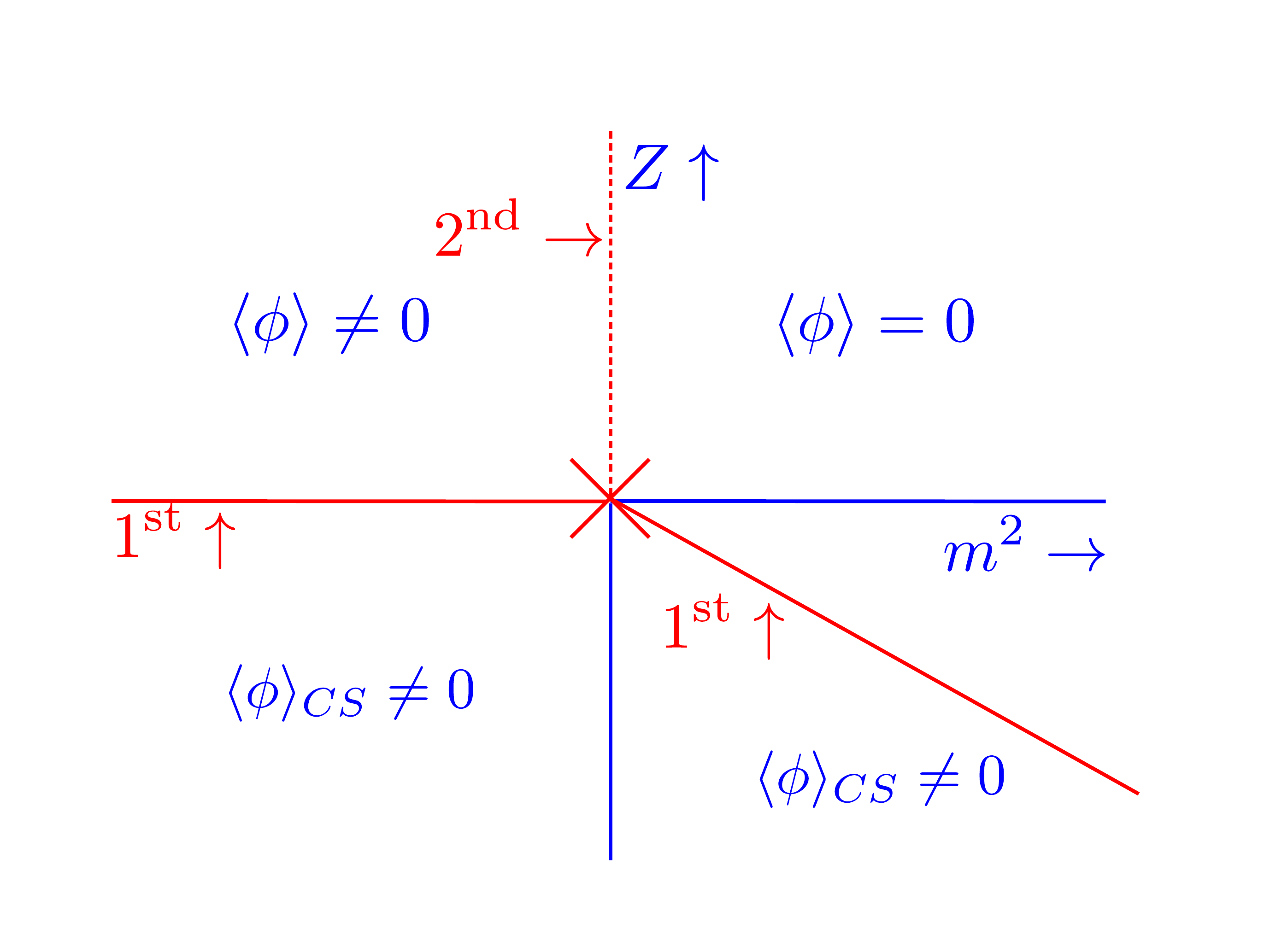}}
  \caption{
    The phase diagram in mean field theory, in the plane of $m^2$ and $\sqz$: a line
    of second order transitions, meeting two lines of first order transitions, which
    meet at the Lifshitz point, where $m^2 = \sqz = 0$.
  }
\label{fig:mean_field}
\end{figure}

This yields the mean field diagram of Fig. \ref{fig:mean_field}.  In
the plane of $m^2$ and $\sqz$, the broken phase with
$\langle \phi \rangle \neq 0$ in the upper left hand quadrant; the symmetric phase,
$\langle \phi \rangle = 0$, in the upper right hand quadrant and part of the lower right
hand quadrant, and the remainder the phase with a chiral spiral.  They meet
at the origin, $m^2 = \sqz = 0$, which is the Lifshitz point.

Analyses of phases with chiral spirals have been carried our in
effective models of QCD, such as the Nambu Jona-Lasinio (NJL)
model.  See, for example, Fig. 6 of Buballa and Carignano \cite{Buballa:2014tba}.
Instead of $m^2$ and $\sqz$, the physical phase diagram is a function of temperature, $T$, and
the baryon (or quark) chemical potential, $\mu$.  In the NJL
model, for the specific interaction assumed, the Lifshitz point coincides with the critical endpoint, but this
is an artifact of the simplest model to one loop order
\footnote{To one loop order, the simplest NJL model involves the determinant
${\rm tr}\log (\slashed{\partial}  + \sigma)$.  This is invariant under a uniform scaling of both $\slashed{\partial}$
and $\sigma$: $\slashed{\partial} \rightarrow \kappa \slashed{\partial}$ and 
and $\sigma \rightarrow \kappa \sigma$.   This implies that the coefficients of $(\partial_\mu \sigma)^2$
and $\sigma^4$ are equal, along with many others.  These relations are no longer valid
once more four fermion couplings are included.  We thank G. Dunne for discussions on this point.}. 

Consider the theory at the Lifshitz point.  The static propagator is
\begin{equation}
  \Delta(\vec{k}) = \frac{{\cal M}^2}{(\vec{k}^2)^2} \; .
  \label{lifshitz_prop}
\end{equation}
At leading order the leading correction to the mass is
\begin{equation}
  \Delta m^2 \sim - \lambda \int d^d k \; \frac{{\cal M}^2 }{(\vec{k}^2)^2 + m^2 {\cal M}^2 } \; .
  \label{lif_mass}
\end{equation}
This develops a logarithmic divergence in the infrared in four dimensions,
which is then the lower critical dimension
\cite{Erzan:1977,Sak:1978,Grest:1978, Bonanno:2014yia,Zappala:2017vjf}.
Corrections to the quartic coupling begin at one loop order as
\begin{equation}
  \Delta \lambda \sim - \lambda^2  \int d^d k \; \frac{{\cal M}^4}{(\vec{k}^2)^2 ((\vec{k}-\vec{p})^2)^2 }\; .
    \label{lif_coupling}
  \end{equation}
This is logarithmically divergent in eight dimensions, which is the upper critical
dimension \cite{Erzan:1977,Sak:1978,Grest:1978}.
This is contrast to an ordinary critical point: for a propagator $\Delta(k) = 1/k^2$,
where the lower and upper critical dimensions are two and four, respectively.

At the Lifshitz point in four spatial dimensions, in the infrared
the logarithmic
divergences always disorder the theory.  This is stronger at nonzero temperature, when $d=3$
and the infrared divergences are power like $\sim 1/m$.
Consequently, once fluctuations are included, there cannot be a true Lifshitz point.

Inhomogenous polymers provide an example of the absence of a Lifshitz point in
three spatial dimensions \cite{Fredrickson:1997,Duchs:2003,Fredrickson:2006,JonesLodge,Cates12}.
The simplest case is a mixture of oil and water.  These separate into droplets of
oil or water, but by adding a surficant to alter the interface tension, other phases emerge.
A related example is a mixture of two different
polymers, formed from monomers of type A and type B.  To this are added A-B diblock copolymers,
which are long sequences of type A, followed by type B.  These A-B copolymers localize at the interfacial
boundaries separating phases with only A or B homopolymers, and act to decrease the interface tension;
at sufficiently high concentrations, the interface tension changes sign, and is negative.

By varying the temperature and the concentration of diblock copolymers one can form
three different phases.  At high temperature A, B, and A-B polymers mingle to form
a homogeneous phase, analogous to the symmetric phase of a spin system.  At low
temperature and low concentrations of A-B copolymers, the system separates into droplets
of A, B and A-B polymers, which is like the broken phase of a spin system.  At low temperature and high concentration
of A-B copolymers, the interface tension becomes negative, and there is an inhomogenous phase, as
the system forms a lamellar state with alternating layers of A and B polymers.
This is similar to a smectic liquid crystal, albeit without orientational order.

Mean field theory predicts that there is a Lifshitz point where these three phases meet.
In contrast, both experiment and numerical simulations with self consistent field theory
indicate that there is {\it no} Lifshitz point \cite{Fredrickson:1997,Duchs:2003,Fredrickson:2006,JonesLodge,Cates12}:
see, {\it e.g.}, Fig. 3 of Ref. \cite{JonesLodge}.
Instead, the symmetric phase enlarges, and includes a bicontinuous microemulsion, which
exhibits nearly isotropic fluctuations in composition with large amplitude.
In this regime the  surface tension is essentially zero, and there is a spongelike structure with large entropy.

The absence of the Lifshitz point can be understood by analogy.
Consider a spin system, with a continuous symmetry, in two or fewer dimensions.
The symmetry cannot be spontaneously broken as that would generate
massless Goldstone bosons, which are not possible in such a low dimensionality.
Instead, fluctuations generate a mass non-perturbatively.

What happens in the Lifshitz regime, when the number of spatial dimensions is four or less,
is similar.  We can tune either the coefficient of the term quadratic in momenta to vanish,
${\cal Z} = 0$, {\it or} the mass, $m^2$, to vanish, but not both.  If $m^2 = 0$, then ${\cal Z} \neq 0$
is generated non-perturbatively; alternately, if ${\cal Z} = 0$, then $m^2 \neq 0$ is generated non-perturbatively.
For the latter, the propagator is not Eq.~(\ref{lifshitz_prop}), but
\begin{equation}
  \Delta(\vec{k}) = \frac{{\cal M}^2 }{(\vec{k}^2)^2 + m^2 {\cal M}^2 } \; ,
  \label{lifshitz_prop_massive}
\end{equation}
where $m^2 \neq 0$ is non-perturbative.
We cannot conclude anything about the size of the Lifshitz regime, only that it exists.
For inhomogeneous polymers, the Lifshitz regime includes
a bicontinuous microemulsion, where ${\cal Z} \approx 0$ and $m^2 \neq 0$;
see, {\it e.g.}, Fig. 2 of Ref.~\cite{JonesLodge}.

\begin{figure}[t]
	\centerline{  \includegraphics[scale=0.5]{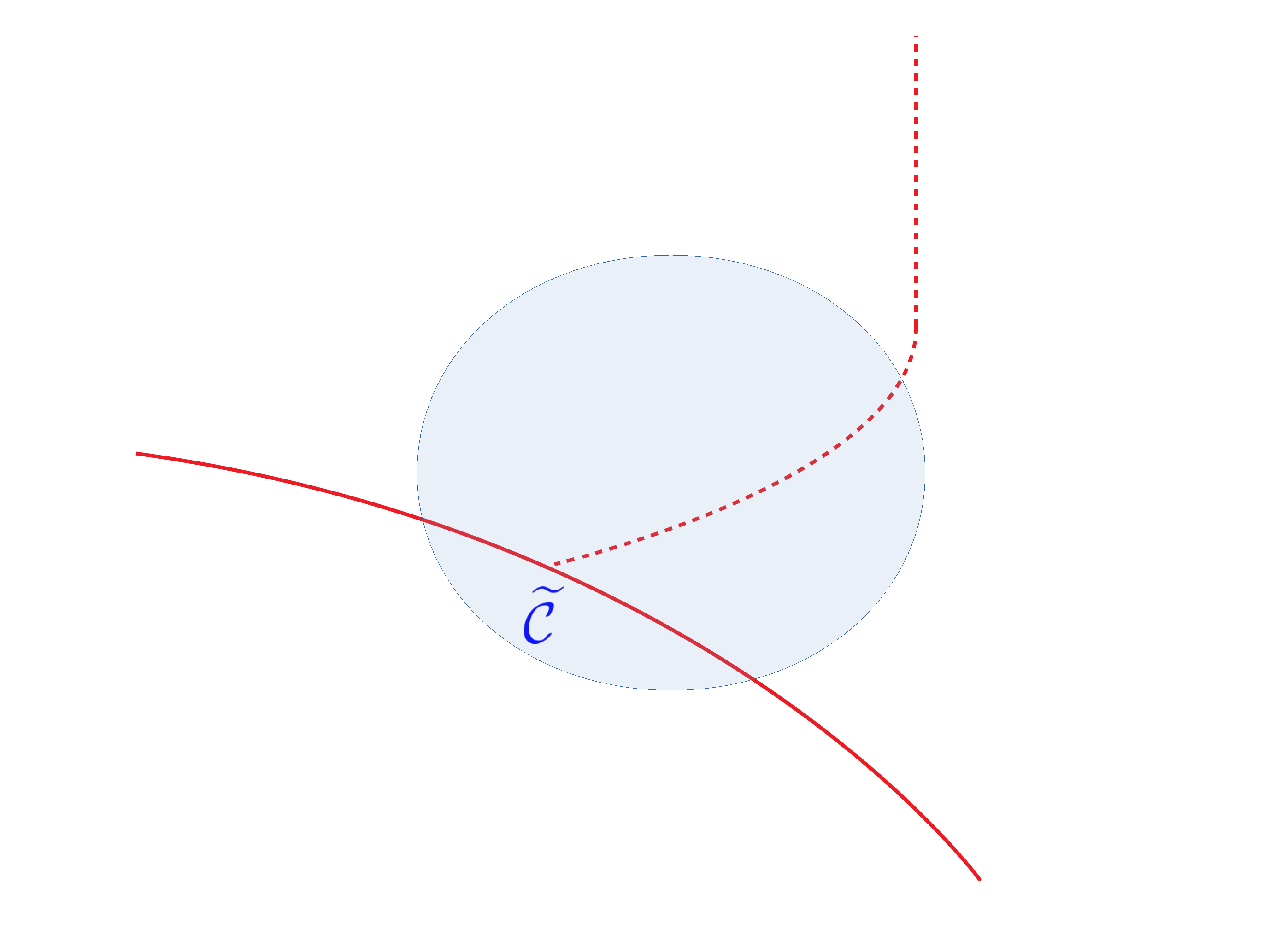}}
  \caption{
    The Lifshitz phase diagram corrected by fluctuations: the line of second order
    transitions still intersects the line of first order transitions, but
    one cannot reach the Lifshitz point, where ${\cal Z} = m^2 = 0$.
    The shaded region denotes the Lifshitz regime, where there are
    large infrared fluctuations.
    The line of second order transitions meets the line of first order transitions at the
    Lifshitz critical endpoint $\widetilde{{\cal C}}$.
  }
\label{fig:fluc_phase}
\end{figure}

A possible phase diagram which incorporates fluctuations is that of Fig. \ref{fig:fluc_phase}.
There is a strict order parameter which distinguishes the broken and symmetric
homogeneous phases, so the line of second order transitions must intersect
the line of first order transitions. They do so at 
a Lifshitz critical endpoint $\widetilde{{\cal C}}$.
By continuity, as $\widetilde{{\cal C}}$ is approached along the line of
first order transitions, the latent heat vanishes.

Consider the usual phase diagram where a line of second order transitions meets a line
of first order transitions at a critical endpoint $\cal C$.  
The universality class along
the line of second order transitions is determined by the unbroken symmetry group and the dimensionality of space,
with nonzero values for the quartic couplings of Eq. (\ref{eff_lag}), $\lambda \neq 0$.
At the critical endpoint $\cal C$, the quartic couplings vanish, $\lambda = 0$, and the
hexatic couplings $\kappa$ dominate.  This changes the
upper critical dimensionality from four to three.

The Lifshitz critical endpoint $\widetilde{{\cal C}}$ is not of this form.
The simplest possibility is that at $\widetilde{{\cal C}}$, a term quadratic in the
momenta, ${\cal Z} > 0$, is generated {\it non}-perturbatively, with $m^2 = 0$.
This implies that the universality class of the Lifshitz critical
endpoint $\widetilde{{\cal C}}$ is the same as along the line of second order transitions.

Consider moving away from the Lifshitz critical endpoint $\widetilde{{\cal C}}$, down
in ${\cal Z}$ into the inhomogeneous phase.
Since mean field theory indicates that an inhomogeneous phase only arises when
${\cal Z}$ is negative, the appearance of an inhomogeneous phase
infintesimally below $\widetilde{{\cal C}}$ must be 
due to strong, non-perturbative fluctuations.

Alternately, consider moving away from the Lifshitz critical endpoint to the right,
for increasing $m^2$.  Doing so, one will enter a region where ${\cal Z}$ is very small,
but the mass squared $m^2$ is nonzero and positive. 
This region is directly analgous to a bicontinuous microemulsion
\cite{Fredrickson:1997,Duchs:2003,Fredrickson:2006,JonesLodge,Cates12}.
For inhomogeneous polymers, this region is seen to be an enlargement of the symmetric
phase into the region between the inhomogenous and broken phases. 
This explains the curvature of the line of second order transitions in Fig. \ref{fig:fluc_phase}.
We do not explicitly indicate the axes ${\cal Z}$ and $m^2$ in Fig. \ref{fig:fluc_phase}
because the Lifshitz point of mean field theory, ${\cal Z} = m^2 = 0$, is not accessible physically.

We note that the phase diagram of mean field theory {\it is} correct in a limit without
fluctuations.  Examples include
Gross-Neveu type models in two spacetime dimensions, which are soluble for
an infinite number of flavors, $N = \infty$
\cite{Schon:2000he,Schnetz:2004vr,Thies:2006ti,Basar:2008im,Basar:2008ki,Basar:2009fg}.
At large but finite $N$, then, the width of the Lifshitz regime is automatically $\sim 1/N$.
It would be useful to study the Lifshitz regime in models with
a large $N$ expansion, both in the lower critical dimension of four and below 
four dimensions.  This would provide a test of
the Lifshitz phase diagram in Fig. \ref{fig:fluc_phase}
and especially of the universality class of the Lifshitz critical endpoint $\widetilde{{\cal C}}$.

Before continuing to the implications for the phase diagram of QCD, we remark
that our analysis is valid for nonzero temperature in three spatial dimensions.
At zero temperature, by causality there must always be terms quadratic in the energy.
The integral analogous to Eq.~(\ref{lif_mass}) then becomes
\begin{equation}
  \Delta m^2 \sim -\lambda \int d\omega \int d^d k \; \frac{1}{\omega^2 + (\vec{k}^2)^2/{\cal M}^2  + m^2}
  \sim -\lambda \int d^d k \frac{{\cal M}}{\sqrt{(\vec{k}^2)^2 + m^2 {\cal M}^2 }} \; .
  \end{equation}
As $m \rightarrow 0$ this is infrared convergent in more than two spatial dimensions, $d > 2$.  Thus we expect that
the infrared fluctuations are well behaved at low temperature.  Further, the dynamic behavior
near the Lifshitz critical endpoint, $\widetilde{{\cal C}}$, differs from
that for a typical critical endpoint, ${\cal C}$.

\section{Relation to QCD}

As we have discussed, the phase diagram 
is a function of at least three parameters: the mass squared, quartic coupling(s), and the 
spatial wave function renormalization ${\cal Z}$.  
At the outset, we assume that the quartic couplings $\lambda_1$ and $\lambda_2$ of
the effective model remain positive, so there is no first order transition associated
with their change of sign.  This assumption can only be decided by numerical simulations
in the underlying theory (which because of the sign problem, is not possible at present), or at least
by using effective theories more closely related to the underlying dynamics.
This qualification needs to be stressed: there could well be {\it both} a critical
endpoint, where quartic coupling(s) $\lambda$ changes sign, {\it and} a line of first order transitions to a spatially
inhomogeneous phase, where ${\cal Z}$ changes sign.

The above analysis applies to the chiral limit, where pions are massless in
the broken phase and there is a line of second order phase transitions.
In QCD, pions are massive in the broken phase, which is similar to having a background field for the chiral
order parameter.  This turns the line of
second order transitions into a crossover line.  Similarly, the
Lifshitz critical endpoint $\widetilde{{\cal C}}$ is also washed out.  We assume that the line of first order
transitions to spatially inhomogeneous phases persists.  

We note that while
their detailed form changes, spatially inhomogeneous phases are relatively insensitive to nonzero
quark masses.  This was explicitly demonstrated in Sec. \ref{model_quarkyonic}, where we treated massive quarks.
Even for heavy quarks, there can be oscillations about a nonzero value for $\overline{q} q$,
as shown by the solution of the 't Hooft model in $1+1$ dimensions \cite{Bringoltz:2009ym}.
Thus the phase with pion/kaon condensates and quarkyonic chiral spirals should perist in QCD.
Further, they are still distinguished by the spontaneous breaking
of a $U(1)$ phase, with associated long range correlations.

\begin{figure}[t]
	\centerline{  \includegraphics[scale=0.5]{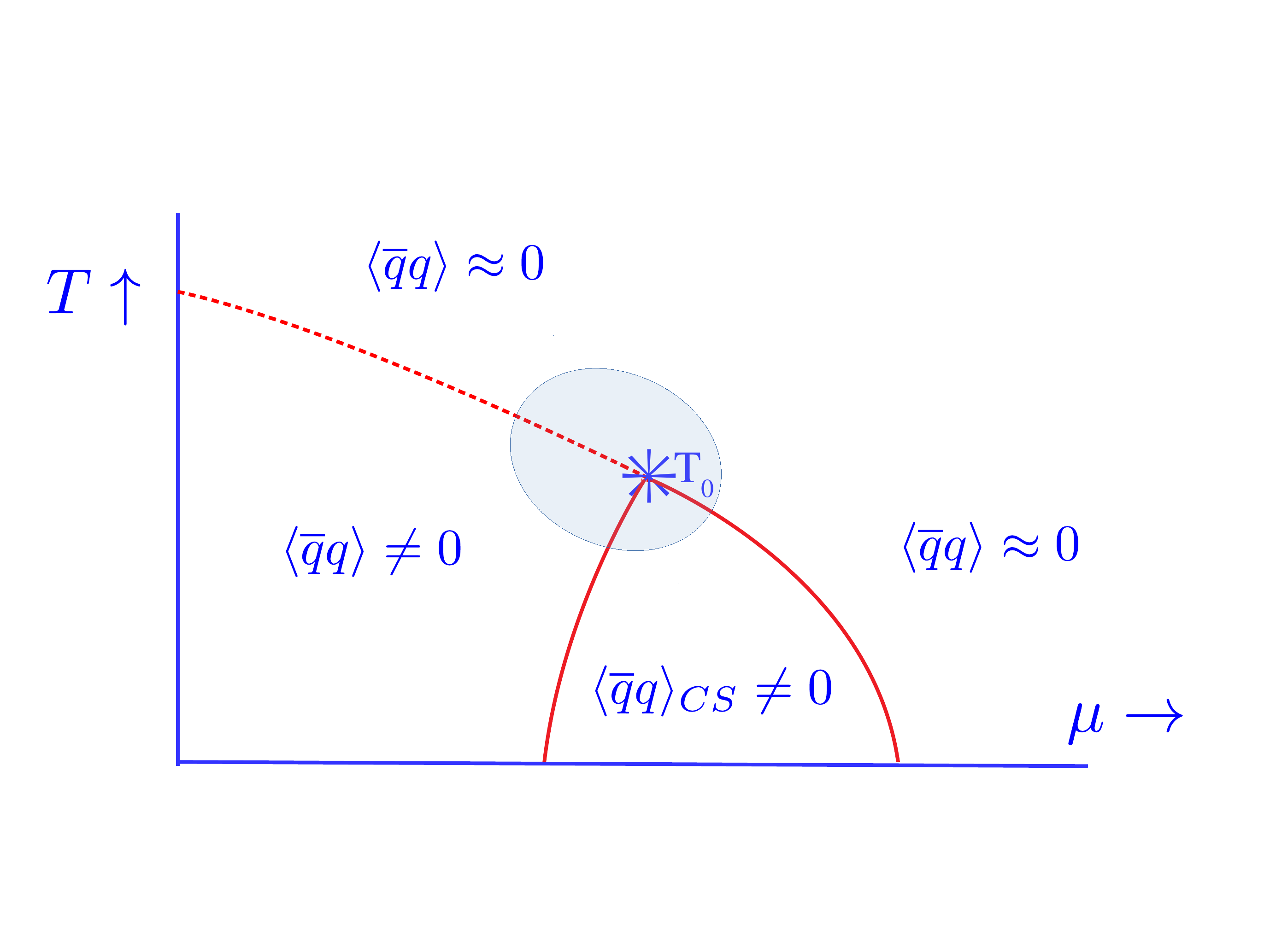}}
  \caption{
    A proposed phase diagram for QCD: the solid line represents first order transitions which
    separate homogeneous from spatially inhomogenous phases; the dashed line, crossover;
    the shaded region, the Lifshitz regime.  The highest
    temperature at which a spatially inhomogeneous phase occurs defines the point of equal densities, $T_0$.  
  }
\label{fig:qcd}
\end{figure}

A caricature of the possible phase diagram in QCD is illustrated in Fig. \ref{fig:qcd}.
The Lifshitz point is wiped out by strong infrared fluctuations, leaving a Lifshitz regime.
We denote this by the shaded regime in Fig. \ref{fig:qcd}, but it is not a precisely
defined region.  The infrared fluctuations in the Lifshitz regime are dominated by massive
modes whose momentum dependence is dominated by quartic terms.

Of particular interest is the highest temperature at which there is a spatially inhomogenous
phase, $T_0$,
\begin{equation}
  \left.  \frac{\partial T}{\partial \mu} \right|_{T_0} = 0 \; .
 \end{equation}
Since the pressure is continuous at a first order phase transition, by taking derivatives
of the pressure with respect to $\mu$, we find
\begin{equation}
  n_+ = \left. \frac{\partial p(T,\mu)}{\partial \mu}\right|_{T_{0}^{-}} =
  n_- = \left. \frac{\partial p(T,\mu)}{\partial \mu}\right|_{T_{0}^{+}} \; .
\end{equation}
This implies that even though there is a first order transition at $T_0$, the densities
are equal.  This is known in thermodynamics as a point of
equal concentration.  Since the transition is of first order, the entropies between
the two phases differ at $T_0$.

We assume that the crossover line terminates at $T_0$, so in the chiral limit,
$T_0$ coincides with the Lifshitz critical endpoint, $\widetilde{{\cal C}}$.  We cannot
prove that $T_0$ is the shadow of $\widetilde{{\cal C}}$, but it is a most natural
conjecture.

What are the possible signals of the phase diagram in Fig. \ref{fig:qcd}?
In heavy ion collisions, assuming that the system thermalizes, it
starts at high temperature and then cools down.  The trajectory
in the plane of temperature $T$, and quark chemical potential, $\mu$, is model dependent, but the
point at which the system freezes out of equilibrium is found by fits to the spectra for
different particle species, and gives values for the final
$T$ and $\mu$.  (The baryon chemical potential is three times that for the quarks.)

The collisions of heavy ions with atomic number $A$ are characterized by the center of mass energy per nucleon.
At the highest energies, $\sqrt{s} = 200$ GeV/A at the 
Relativistic Heavy Ion Collider (RHIC), and $\sqrt{s} = 3$~TeV/A at the
Large Hadron Collider (LHC), the quark chemical potential at freezeout is small.
At lower energies, $\sqrt{s}/A: 1 \rightarrow 20$~GeV, 
fits to thermal models
\cite{Andronic:2009gj,Andronic:2014zha,Andronic:2017pug,Adamczyk:2017iwn,Bugaev:2017nir,Bugaev:2017bdc,Bugaev:2018tyt}
demonstrate that one enters a region where the quark chemical potential at freezeout is significant.

The standard picture 
\cite{Asakawa:1989bq,Stephanov:1998dy,Stephanov:1999zu,Son:2004iv,Hatta:2002sj,Stephanov:2008qz}
assumes the crossover line for small $\mu$ meets a line of first order transitions at a critical endpoint as
$\mu$ increases and $T$ decreases.  At a critical endpoint, in infinite
volume and in thermal equilibrium, there are divergent fluctuations for 
the critical mode, which is associated with the $\sigma$ meson.  There should also be large
fluctuations for modes which couple to the $\sigma$ meson, including pions, kaons, and nucleons.
It is not possible to measure the fluctuations for $\sigma$'s directly, but as we discuss below,
it is possible to measure that for protons.  Measuring the fluctuations for pions and kaons
is experimentally very challenging, but we argue is essential in order to distinguish between
different models.  As the lighter particle, near a critical endpoint ${\cal C}$ the fluctuations for pions
should be greater than for kaons.

{\it If} the phase diagram does not have a critical endpoint, but instead has an unbroken
line of first order transitions as in Fig. \ref{fig:qcd}, then the signals depend upon
the trajectory in the $T-\mu$ plane.  One obvious difference is that with Fig. \ref{fig:qcd},
it is possible to cross two first order lines before hadronization occurs.

We first discuss the case in which the system enters the Lifshitz regime but is still
in the symmetric phase, before it crosses the line of first order transitions.
In principle it is necessary to include a nonzero chemical potential for up and down quarks
and to impose the condition that the net strangeness vanishes.  We leave these details to
future study to make the following elementary point.

Consider a particle with the usual dispersion relation,
$E = \sqrt{\vec{k}^2 + m^2}$.  In the limits of high and low temperature the average momentum is
\begin{equation}
  T \gg m: \;\;\; \langle k \rangle \sim T \;\;\; ; \;\;\;
  T \ll m: \langle k \rangle \sim \sqrt{m \, T} \; .
  \label{usual_avg_mom}
\end{equation}
In the ultra-relativistic limit the average momentum is necessarily independent of
mass and is proportional to the only mass scale, which is the temperature.  In the non-relativistic limit
the average momentum is proportional to the square root of the mass, times the temperature.

Now consider a particle in the Lifshitz regime, assuming that the coefficient of the
term quadratic in the spatial momentum is essentially zero.  The dispersion relation is
then
\begin{equation}
  E_{\rm Lifshitz} = \sqrt{ \frac{(\vec{k}^2)^2}{M^2} + m^2} \; .
\label{Lifshitz_dispersion}
\end{equation}
The mass scale $M$ ensures that the term quartic in the spatial momentum, Eq. (\ref{braz_prop}),
has the correct mass dimension.
As discussed above, this dispersion relation is analogous to the bicontinuous microemulsion phase of
inhomogenous polymers \cite{Fredrickson:1997,Duchs:2003,Fredrickson:2006,JonesLodge,Cates12}.
For such a dispersion relation, the average momentum in the limits of high and low temperature is
\begin{equation}
  T \gg m: \;\;\; \langle k \rangle_{\rm Lifshitz} \sim \sqrt{ M \, T} \;\;\; ; \;\;\;
  T \ll m: \langle k \rangle_{\rm Lifshitz} \sim (m \, M^2 \, T)^{1/4} \; .
  \label{lifshitz_avg_mom}
\end{equation}
Again, in the ultra-relativistic limit the average momentum is independent of the mass $m$,
but now it is only proportional to the square root of temperature, with $M$ making up the remaining
mass scale.  In the limit of low temperature, the average momentum is proportional
not to the square root of the mass, but to the fourth root thereof.

In heavy ion collisions, the freezeout temperature is near the pion mass, so for simplicity
we assume that the pions are ultra-relativistic.  For
kaons, we assume that they are non-relativistic.  Of course this is a gross simplification,
but it not difficult to carry out a more careful analysis in a thermal model.

Because the dispersion relation in the Lifshitz regime
differs fundamentally from the usual relation, the relative
abundance of kaons to pions {\it must} change when these particles are in the Lifshitz regime.
In particular, the mass dependence for heavy particles, such as kaons, is less sensitive
to mass, $\langle k \rangle \sim m^{1/4}$, Eq. (\ref{lifshitz_avg_mom}), versus
$\langle k \rangle \sim m^{1/2}$ in Eq. (\ref{lifshitz_avg_mom}).
Thus in the Lifshtiz regime, the ratio of kaons to pions is {\it greater} than a fit with
a standard thermal model.

The difference between pions and kaons persists
once spatially inhomogeneous condensates develop.  In the 
simplified discussion of Sec. (\ref{general_phase}), we did not distinguish between
pions and kaons.  This valid in the strict chiral limit, but not in QCD.
Moving down in temperature at fixed $\mu$, presumably a pion condensate develops
before that for kaons.  Indeed, pion condensates are naturally chiral spirals,
rotating between $\sigma$ and a given direction for the pions.  In contrast,
kaons presumably develop a kink crystal first, oscillating about 
a given expectation value for $\overline{s} s$.
As the chemical potential increases at a fixed, small value of
the temperature, these condensates then evolve into a chiral
spiral of the quarkyonic phase, and approach the $SU(3)$ symmetric limit
in flavor.  In any case, the effective masses for fluctuations are given by
Eq. (\ref{eff_mass}), and differ markedly from those of free particles.

For each particle species, in a phase with spatially inhomogenous condensates
the fluctuations are concentrated {\it not} about zero momentum
but about the momentum for the condensate, $Q_0$ in Eq. (\ref{k0_min}).  This should
be measurable by measuring the fluctuations in different bins in momenta.
This is challenging experimentally, as any condensate is with respect to the
local rest frame, which is boosted by hydrodynamic expansion to
a significant fraction of the speed of light.

Depending upon 
the trajectory in the plane of $T$ and $\mu$, it may be possible to cross not
just one, but two lines of first order transitions before the
system hadronizes.  Lastly, the point $T_0$
is of especial interest, although it is not clear whether trajectories
naturally flow into it.

Before discussing heavy ion experiments, we note that Andronic {\it et al.}
\cite{Andronic:2009gj} argued that there is a triple point in the $T-\mu$ plane.
The Lifshitz regime can be considered as an explicit way of generating this phenomenon.

There are two notable anomalies in the collisions of heavy ion at relatively
low energies.

The first is a strong departure from thermal models.
In heavy ion collisions, there is a peak
in the ratio of $K^+/\pi^+$ and $\Lambda/\pi$ at energies $\sim 10$~GeV
\cite{Andronic:2009gj,Andronic:2014zha,Andronic:2017pug,Adamczyk:2017iwn,Bugaev:2017nir,Bugaev:2017bdc,Bugaev:2018tyt}.
Deviations from thermal behavior for the ratio of kaons to pions is suggestive of the Lifshitz
regime, Eqs. (\ref{usual_avg_mom}) and (\ref{lifshitz_avg_mom}) above.
However, the ratio $K^-/\pi^-$ shows no such deviation.  Clearly a more careful analysis, including
the condition of zero net strangeness, is essential.  

The second anomaly concerns fluctuations in net protons.  Experimentally it is
possible to measure cumulants, which are
related to the derivatives of the pressure with respect to the chemical potential,
\begin{equation}
  c_n(T, \mu) \sim \frac{\partial}{\partial \mu^n} \; p(T,\mu)  \; .
\label{define_moments}
\end{equation}
The results from numerical simulations on the lattice appears to agree
remarkably well with the predictions of lattice gauge theory except at
the lowest energies
\cite{Bellwied:2015lba,Bellwied:2015rza,Gunther:2016vcp,Bazavov:2017tot,Bazavov:2017dus,Vovchenko:2017gkg,Almasi:2018lok}.
There, unpublished data from the Beam Energy Scan with the STAR experiment at RHIC suggests a possible
anomaly at $\sqrt{s}/A \sim 8$~GeV \cite{Jowzaee:2017ytp,Luo:2017faz}.

For the ratio of the fourth to the second cumulant, $c_4/c_2$, when only net protons
with transverse momenta between $0.4$ and $0.8$~GeV
are included, this ratio is essentially flat from the highest energy, $\sqrt{s}/A = 200$ GeV, down to the
lowest, $\sim 8$~GeV.  It is one above $40$ GeV, then decreases to $\sim 0.6 - 0.8$ below $40$ GeV, with
large error bars at the lowest energies.

However, if net protons with transverse momenta
between $0.4$ and $2.0$ GeV are included, the ratio, again with large error bars,
shows striking non-monotonic behavior, decreasing from one at high energy, to $\sim 0.3 \pm 2$ at
$20-30$ GeV/A, and then rises sharply, reaching $c_4/c_2 \sim 3.5$ at the lowest energy.

In the Lifshitz regime, pions and kaons behave strongly non-pertubatively, and this feeds
into the fluctuations of protons.  This could explain this possible anomaly.

To distinguish between a Lifshitz regime and a critical endpoint, 
it is essential to measure the fluctuations of
pions and kaons.
Near a critical endpoint the fluctuations for pions are larger
than for kaons.  This is not true in the Lifshitz regime, either in the symmetric or
chiral spiral phase.  It is very possible that for a certain range of energies, that
the fluctuations for kaons exceed those for pions.

Of course evidence for crossing not just one, but two phase transitions, would also be
exceptional.  Nevertheless, without model calculations we cannot estimate how strong the
first order transitions are.  

In conclusion, it is surprising that there are such close analogies between the phase transitions in
condensed matter systems, such as smectics and inhomogenous polymers, and those of QCD.
While our analysis is a first step, it may directly impact our understanding of
the collisions of heavy ions at low energies.

\acknowledgments
R.D.P. thanks G. Dunne for discussions on NJL models, 
the organizers of the Seventh International Conference on
New Frontiers in Physics for the invitation to speak on this work,
and K. Bugaev and K. Redlich for discussions at this meeting.
R.D.P. is funded by the U.S. Department of Energy for support
under contract DE-SC0012704; A. M. T. is funded by Condensed Matter Physics and
Materials Science Division, under the the U.S.
Department of Energy, contract No. DE-SC0012704.

\bibliography{lifshitz}

\end{document}